\documentclass[letterpaper,twocolumn,10pt]{article}
\usepackage{usenix,epsfig}
\usepackage[hyphens]{url}
\usepackage{hyperref}
\usepackage[utf8]{inputenc}
\usepackage{array}
\usepackage[usenames]{xcolor}
\begin{document}

\date{}

\title{\Large \bf Censors' Delay in Blocking Circumvention Proxies}

\author{
{\rm David Fifield\thanks{Authors are listed in alphabetical order.}}\\
University of California, Berkeley
\and
{\rm Lynn Tsai}\\
University of California, Berkeley
}

\maketitle


\subsection*{Abstract}

Censors of the Internet must continually discover and block
new circumvention proxy servers.
We seek to understand this process; specifically,
the length of the delay between when a proxy first becomes
discoverable and when it is actually blocked.
We measure this delay in the case of obfuscated Tor bridges,
by testing their reachability
before and after their introduction into Tor Browser.
We test from sites in the U.S., China, and Iran,
over a period of five months.
China's national firewall blocked new bridges
after a varying delay of between 2 and 36 days.
Blocking occurred only after end-user software releases,
despite bridges being potentially discoverable earlier through other channels.
While the firewall eventually discovered the bridges of Tor Browser,
those that appeared only in Orbot,
a version of Tor for mobile devices, remained unblocked.
Our findings highlight the fact that censors can behave
in ways that defy intuition,
presenting difficulties for threat modeling but
also opportunities for evasion.

\section{Introduction}

Censors of the Internet must seek out and block
proxy servers that can be used to evade their information controls.
Here we explore the mechanics of this process as it applies
to the blocking of default Tor bridges
after they are published in Tor Browser.
It is known that these default bridges are eventually blocked;
what is not known is exactly how long it takes.
We measure the ``delay'' or ``lag'' of proxy blocking,
by testing the reachability of bridges
before and after their first public disclosure
from sites in the U.S., China, and Iran.

There is prior work on distribution strategies that
prevent the censor from discovering secret proxy addresses in the first place;
examples are
Proximax~\cite{McCoy2011a} and
rBridge~\cite{Wang2013a}.
Our work is different: we study how censors block proxies
that are not secret, but are (in principle) easily discoverable by anyone.
In typical censorship threat models,
such unprotected proxies would be considered to be immediately blocked.
That they are not, in practice,
shows that practical considerations
may prevent censors from exercising all their assumed capabilities.


Zhu et~al.~\cite{Zhu2013a} in 2013 explored a related idea.
They measured how long it took for posts to be censored
on the Chinese microblogging service Sina Weibo.
They found that 30\% of posts are deleted within 30~minutes
and 90\% are deleted within 24~hours, though some posts
survived for weeks or months.
They used their results to hypothesize about the reasons for
and mechanisms of microblog censorship.

In 2014, developers of OONI, the Open Observatory of Network Interference,
a censorship measurement platform, implemented a new test to
check the reachability of Tor bridges
and prepared visualizations~\cite{ooni-bridge-reachability-study-and-hackfest}.
The tests have not run continually since then.
In 2015, a one-off calculation~\cite{tor-dev-censorship-lag}
based on user reports
found loose time bounds for the delay of a single bridge-blocking event in China:
somewhere between 15 and 76 days.

Knowing the proxy blocking delay gives insight
into how censors work:
where they look for new bridges, and
whether their blacklist updates are automatic or manual.
It advances our understanding the operational costs incurred by censors,
and therefore their potential weaknesses.
We think of censors as complex systems,
consisting of interacting human and machine
components, whose goals and
motivations are sometimes in conflict.

\section{Background}

Tor~\cite{tor} is an anonymity network
that is also widely used to circumvent censorship.
In its natural form, Tor is poorly suited to circumvention.
Its nodes' IP addresses are public,
and the protocol itself is fairly distinctive.
But Tor combined with \emph{bridges}
and \emph{pluggable transports} is much harder to block.
Bridges are special, unlisted nodes
whose addresses are not easily discoverable in bulk---users
must acquire a few at a time through an online database called BridgeDB~\cite{bridgedb}.
Pluggable transports
are obfuscation protocols that encapsulate the Tor protocol,
making it difficult to detect.
A censored user uses a pluggable transport
in order to reach a secret bridge,
foiling both deep packet inspection
and IP address blacklisting.

Tor Browser~\cite{torbrowser} is a modified version of Firefox
that features a built-in always-on Tor client.
It is the recommended way of accessing the Tor network for most users.
The browser has a graphical interface
for the configuration of bridges and pluggable transports.
The intended use case has
users acquiring a bridge address through a side channel,
such as email or word of mouth,
and pasting it into the configuration interface.
However, in practice, circumvention is often even easier,
requiring no out-of-band information.
For many users, it suffices to select a pluggable transport name from a menu,
causing Tor to connect to one of
a handful of built-in, \emph{default} bridges
included with the browser.

The concept of a default bridge needs some explanation.
Bridges are supposed to be secret,
so including them with the browser seems to be self-defeating.
The default bridges are in fact open to the world
in the Tor Browser source code---the file housing them is called
bridge\_prefs.js~\cite{torbrowser-bridgeprefs}.
Any censor that can block Tor ought to
be able to block the default bridges as well,
but the strange truth of the matter is that many simply do not.
We can only guess as to why:
it could be that censors are negligent or incompetent;
perhaps Tor traffic is, from their point of view, too inconsequential to bother with;
or maybe they set more stock in deep packet inspection
and dynamic protocol detection than in static IP address blacklists.
Whatever the reason, Tor Browser continues to ship
default bridges for the simple reason that they work for many people.
In fact, the Great Firewall of China is the only state censor
we are aware of that makes some effort to block Tor's default bridges.
This is what we study in this work:
the time delay in blocking a bridge after it is first made public
in a Tor Browser release.

\subsection{Tor Browser releases}

Tor Browser releases follow two tracks: stable and alpha.
The stable track changes slowly,
its minor releases typically including only bugfixes.
The alpha track has new and experimental features.
The alpha track matures until it becomes the basis
of the next major release of the stable track.
(There are also special ``hardened'' releases
that track the alphas---we do not consider them separately.)
In addition to formal major and minor releases,
there are nightly builds that have the latest of everything.

The primary driver of Tor Browser releases is
upstream security fixes in Firefox.
When there is a new version of Firefox
that fixes a security vulnerability,
Tor Browser developers must hurry to
build a new minor release of Tor Browser with the fix.
For this reason, new stable and alpha releases
usually appear at about the same time,
because both are usually
equally affected by Firefox bugs.
Alpha releases are distinguished by a letter ``a''
in the version number;
for example, the stable release 5.5.5
appeared at about the same time as
the alpha release 6.0a5.
During the roughly five-month period of our experiments,
there were
ten stable releases (two major, eight minor), and
seven alpha releases (one major, six minor)~\cite{torbrowser-changelog}.
Each releases was an opportunity to
deploy and measure new bridges.

\subsection{The process of releasing a new bridge}
\label{sec:bridge-process}

There is no single moment when a new bridge becomes public.
The process of adding a bridge involves multiple stages,
each of which potentially reveals it,
depending on how closely the censor pays attention.

\begin{enumerate}
\item \textbf{Ticket filed.}
New default bridges are proposed for inclusion
by the filing of a ticket in Tor's
online bug tracker.
A censor paying attention to the bug tracker
would learn of bridges at this stage.
\item \textbf{Ticket merged.}
When the ticket is merged,
the new bridge is added to Tor Browser's source code.
From there, it will automatically be incorporated
into nightly builds. This is the first time that the code
containing the new bridge is available in executable form.
A censor paying attention to the source code repository,
or following nightly builds,
would learn of bridges at this stage.
\item \textbf{Testing release.}
When it is time for a new release,
Tor Browser developers prepare candidate packages
and send them to a quality assurance mailing list
for testing.
A censor paying attention to the mailing list
would learn of bridges at this stage.
\item \textbf{Public release.}
After testing, new packages are announced on the Tor blog,
and already installed Tor Browsers will automatically update themselves.
This is the stage at which the new bridge
will start to be used by real users.
A censor paying attention to the blog,
or black-box testing an auto-updating installation,
would learn of bridges at this stage.
\end{enumerate}

From ticket filing to public release, the process usually takes a few weeks.
Sometimes the testing release stage is skipped if the new version
only fixes a small error in the previous version, like
a packaging or localization bug.
The releases
of the stable and alpha tracks are separate; they are, however,
usually close in time.

New default bridges are sometimes discussed in private mailing lists
even before a ticket is filed.
A censor could conceivably infiltrate an internal mailing list
and learn about new bridges very early.
We have assumed that this does not happen,
that censors must use the same public channels as everyone else.

\subsection{The special properties of obfs4}

\begin{figure*}
\includegraphics{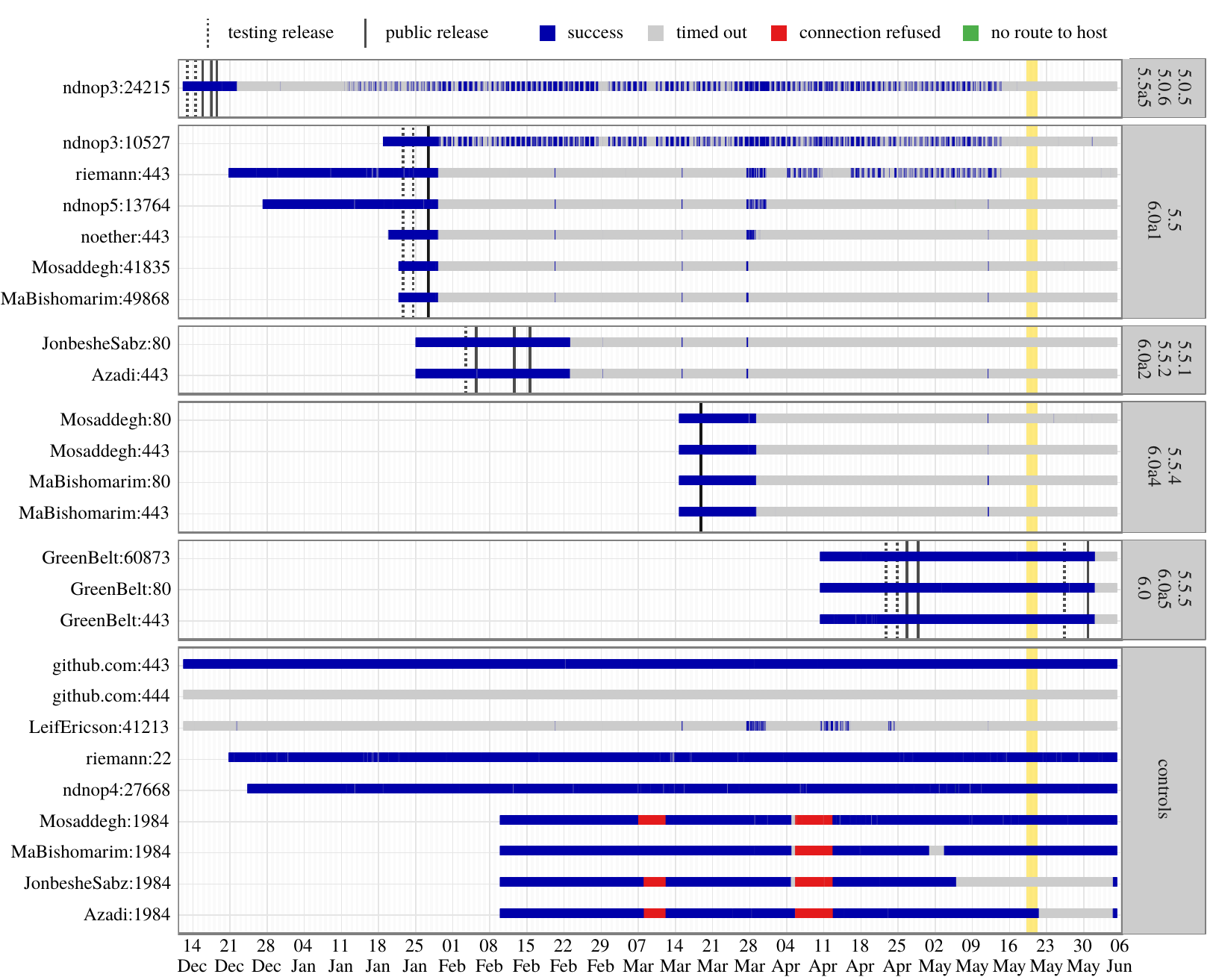}
\caption{
Timeline of Tor Browser default bridge blocking,
as measured from one probe site in China.
Black vertical lines indicate releases.
All new bridges within a release are blocked
within a few days or weeks.
The ``timed out'' and ``connection refused'' results
for the control bridges on port 1984
were the result of temporary misconfigurations, not blocking.
The vertical stripe on May~20 shows when the probe site
changed its IP address.
}
\label{fig:timelines}
\end{figure*}

\begin{figure*}
\includegraphics{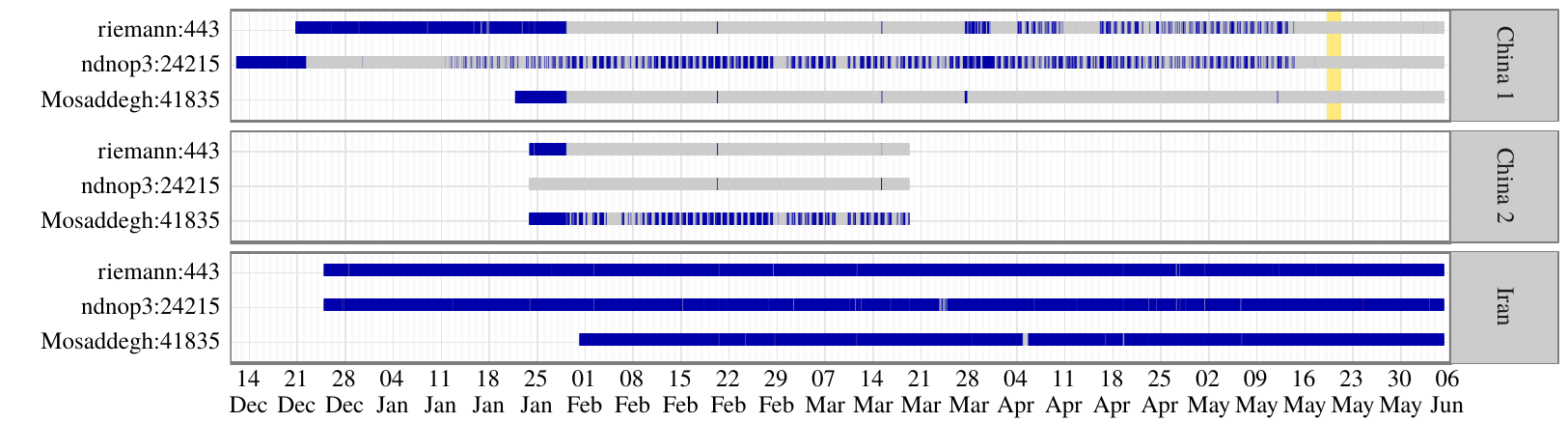}
\caption{
Comparison of reachability of selected bridges
across our three experimental sites.
We detected no blocking from our Iran site.
In China, blocking manifested as timed-out connections.
Blocking in our two China sites was the same,
except for the curious diurnal semi-blocking pattern,
which affected different bridges at each site.
}
\label{fig:comparison}
\end{figure*}

Though Tor Browser supports several pluggable transports,
we concern ourselves only with obfs4~\cite{obfs4},
an advanced transport offering several security features.
We rely critically on these features;
they enable us to limit the means of learning
about new bridges to the ones we control.
The use of obfs4 gives us confidence that the censor
learns of our bridges only in the ways we intend,
namely their inclusion in Tor Browser.

obfs4 resists deep packet inspection by re-encrypting
a Tor stream so that it appears as a stream of random bytes.
More than that, obfs4 resists \emph{active probing} attacks
in which the censor scans suspected proxies
in order to discover what protocols they support.
Every obfs4 server has a per-bridge secret,
which the client must prove knowledge of in its initial message.
The Great Firewall is known to employ active probing against the predecessor protocols
obfs2 and obfs3~\cite{Ensafi2015b},
but the same attack is ineffective against obfs4.
The censor must have the same out-of-band
information as a legitimate client;
merely knowing the IP address of a bridge is insufficient
to confirm that it is, in fact, a bridge.


In addition to its useful properties, obfs4 is relevant to real-world users.
It is marked ``recommended'' in Tor Browser
and has the more users than any other transport,
about 20,000 concurrent on average as of June~2016~\cite{obfs4-users}.
obfs4 has experimental support for obfuscating
packet size and timing,
but none of the default bridges have deployed that feature,
so we did not investigate it.

\section{Method}

We established probe sites in three countries,
using dedicated servers or cloud services:
one in the U.S., two in China, and one in Iran.
The two sites in China were in different autonomous systems.
At each site, we ran a script that tested the TCP reachability
of a variety of destinations every 20 minutes.
For each destination, the script attempted to establish a TCP connection
and then recorded the current time, IP address, port number, status,
and error message if any.
We ran probes for about five months,
from December~12, 2015 to June~4, 2016.
The destinations we tested,
a mixture of fresh obfs4 bridges and other control destinations,
appear in Table~\ref{tab:bridges}.

The probe site in the U.S. was a control
that enabled us to distinguish cases of censorship
from a bridge's temporarily being down.
Owing to the difficulty of accessing network services in China and Iran,
the time periods during which we had access to each site
differ, though they overlap.
In one of the China sites, we lost access to our probing host
partway through the experiment; however, before that happened
we got access to another in the same autonomous system.
The data in Figure~\ref{fig:timelines} and Table~\ref{tab:timeline},
which show the results from this site,
are spliced together from the two series.

\begin{table}
\begin{tabular}{l l}
nickname & ports \\
\hline
\noalign{\smallskip\normalsize\textbf{New Tor Browser default bridges}}
ndnop3 & 24215, 10527 \\
ndnop5 & 13764 \\
riemann & 443 \\
noether & 443 \\
Mosaddegh & 41835, 443, 80 \\
MaBishomarim & 49868, 443, 80 \\
JonbesheSabz & 80 \\
Azadi & 443 \\
GreenBelt & 60873, 80, 443 \\
\noalign{\smallskip\normalsize\textbf{Orbot-only default bridges}}
Mosaddegh & 1984 \\
MaBishomarim & 1984 \\
JonbesheSabz & 1984 \\
Azadi & 1984 \\
\noalign{\smallskip\normalsize\textbf{Already existing Tor Browser default bridges}}
LeifEricson & 41213 \\
\noalign{\smallskip\normalsize\textbf{Never-published bridges}}
ndnop4 & 27668 \\
\noalign{\smallskip\normalsize\textbf{Non-bridge controls}}
github.com & 443, 444 \\
\end{tabular}
\caption{
The destinations, mostly obfs4 bridges, whose reachability we tested.
Bridges are identified by their ``nickname,'' an arbitrary label assigned by the bridge's operator.
Each nickname represents a distinct IP address;
some bridges served obfs4 on multiple ports.
In addition to the destinations listed,
we tested
port 22 (SSH) on the bridges that had it open.
}
\label{tab:bridges}
\end{table}

We were fortunate to run this experiment at a time
when Tor Browser was ramping up its obfs4 capacity,
with new bridges being added in nearly every release.
In some cases these were new ports on existing IP addresses;
in others they were entirely new IP addresses.
We began measurements of each new bridge as soon as we became aware of it.
In some cases we received advance notice of a new bridge
before its ticket was filed,
but in others we started measurements shortly after the ticket.
We coordinated with the Tor Browser developers
to ensure that newly created bridges were not present
in BridgeDB~\cite{bridgedb}, where they might have been
discovered by censors and ordinary users.
We tried, as far as possible, to limit the possible
avenues of discovery to bug tracker tickets and the Tor Browser source code.

We also measured four bridges that appeared only in Orbot~\cite{orbot},
the port of Tor to Android.
Orbot and Tor Browser have most of their default bridges in common,
but a few appear in Orbot only.
They are the ones with port number 1984 in Table~\ref{tab:bridges}.
The Orbot-only bridges remained accessible,
even as the Tor Browser bridges were blocked.

\section{Results}

We recorded over 1.5~million individual probe results
over a period of approximately five months.
The results from our two China sites were the same in most respects,
with blocking occurring at the same time in both.
We found no blocking at all of the default bridges from our site in Iran.
As the two China sites were similar and the Iran site did not show any blocking,
we will mainly present the results from the China site that had more data
(called ``China~1'' in figures).
A graphical summary of the results appears in
Figure~\ref{fig:timelines}
and a textual timeline in Table~\ref{tab:timeline}.
Figure~\ref{fig:comparison} compares
the three sites across a subset of destinations.

\newcommand{\eventheading}[1]{\textbf{#1}\quad}
\newcommand{\ticketfiled}{\eventheading{Ticket filed}}
\newcommand{\ticketmerged}{\eventheading{Ticket merged}}
\newcommand{\testrelease}{\eventheading{Testing release}}
\newcommand{\pubrelease}{\eventheading{Public release}}
\newcommand{\blocked}{\eventheading{Blocked}}

\newcommand{\releaseheading}[2]{\noalign{\smallskip\normalsize\textbf{#1:} #2}}
\newcommand{\releaseheadingnull}[1]{\noalign{\smallskip\normalsize\textcolor{gray}{\textbf{#1:} no new bridges}}}

\begin{table}
\footnotesize
\setlength{\tabcolsep}{2pt}
\begin{tabular}{l r >{\raggedright\arraybackslash}p{2.2in}}
\releaseheading{Tor Browser 5.0.5/5.0.6/5.5a5}{1 new bridge}
03 Dec & $-$19 days & \ticketfiled  ndnop3:24215 \\
07 Dec & $-$15 days & \ticketmerged ndnop3:24215 \\
12 Dec & $-$10 days & \testrelease  5.0.5 stable \\
14 Dec &  $-$8 days & \testrelease  5.5a5 alpha \\
15 Dec &  $-$7 days & \pubrelease   5.0.5 stable \\
17 Dec &  $-$5 days & \pubrelease   5.0.6 stable \\
18 Dec &  $-$4 days & \pubrelease   5.5a5 alpha \\
22 Dec & 0\phantom{ days} & \blocked      ndnop3:24215 \\
\releaseheadingnull{Tor Browser 5.0.7/5.5a6}
\releaseheading{Tor Browser 5.5/6.0a1}{6 new bridges}
16 Jan & $-$13 days & \ticketfiled  riemann:443 \\
18 Jan & $-$11 days & \ticketmerged riemann:443 \\
18 Jan & $-$11 days & \ticketfiled  ndnop3:10527, ndnop5:13764 \\
18 Jan & $-$11 days & \ticketmerged ndnop3:10527, ndnop5:13764 \\
19 Jan & $-$10 days & \ticketfiled  noether:443 \\
19 Jan & $-$10 days & \ticketmerged noether:443 \\
21 Jan &  $-$8 days & \ticketfiled  Mosaddegh:41835, MaBishomarim:49868 \\
21 Jan &  $-$8 days & \ticketmerged Mosaddegh:41835, MaBishomarim:49868 \ \\
22 Jan &  $-$7 days & \testrelease  5.5 stable \\
24 Jan &  $-$5 days & \testrelease  6.0a1 alpha \\
27 Jan &  $-$2 days & \pubrelease   5.5 stable \\
27 Jan &  $-$2 days & \pubrelease   6.0a1 alpha \\
29 Jan & 0\phantom{ days} & \blocked      ndnop3:10527, riemann:443, ndnop5:13764, noether:443, Mosaddegh:41835, MaBishomarim:49868 \\
\releaseheading{Tor Browser 5.5.1/5.5.2/6.0a2}{2 new bridges}
24 Jan & $-$30 days & \ticketfiled JonbesheSabz:80, Azadi:443 \\
28 Jan & $-$26 days & \ticketmerged JonbesheSabz:80, Azadi:443  \\
03 Feb & $-$20 days & \testrelease 5.5.1 stable \\
05 Feb & $-$18 days & \pubrelease 5.5.1 stable \\
12 Feb & $-$11 days & \pubrelease 5.5.2 stable \\
15 Feb &  $-$8 days & \pubrelease 6.0a2 alpha \\
23 Feb & 0\phantom{ days} & \blocked JonbesheSabz:80, Azadi:443 \\
\releaseheadingnull{Tor Browser 5.5.3/6.0a3}
\releaseheading{Tor Browser 5.5.4/6.0a4}{4 new bridges}
12 Mar & $-$17 days & \ticketfiled Mosaddegh:80, Mosaddegh:443, MaBishomarim:80, MaBishomarim:443  \\
14 Mar & $-$15 days & \ticketmerged Mosaddegh:80, Mosaddegh:443, MaBishomarim:80, MaBishomarim:443 \\
18 Mar & $-$11 days & \pubrelease 5.5.4 stable \\
18 Mar & $-11$ days & \pubrelease 6.0a4 alpha \\
29 Mar & 0\phantom{ days} & \blocked Mosaddegh:80, Mosaddegh:443, MaBishomarim:80, MaBishomarim:443 \\
\releaseheading{Tor Browser 5.5.5/6.0a5/6.0}{3 new bridges}
05 Apr & $-$57 days & \ticketfiled GreenBelt:60873, GreenBelt:80, GreenBelt:443 \\
07 Apr & $-$55 days & \ticketmerged GreenBelt:60873, GreenBelt:80, GreenBelt:443 \\
22 Apr & $-$40 days & \testrelease 5.5.5 stable \\
24 Apr & $-$38 days & \testrelease 6.0a5 alpha \\
26 Apr & $-$36 days & \pubrelease 5.5.5 stable \\
28 Apr & $-$34 days & \pubrelease 6.0a5 alpha \\
26 May &  $-$6 days & \testrelease 6.0 stable \\
30 May &  $-$2 days & \pubrelease 6.0 stable \\
01 Jun & 0\phantom{ days} & \blocked GreenBelt:60873, GreenBelt:80, GreenBelt:443 \\

\end{tabular}
\caption{
Timeline of Tor Browser releases
and bridge blocking.
For the meaning of terms, see Section~\ref{sec:bridge-process}.
Time offsets are given from the date of bridge blocking.
}
\label{tab:timeline}
\end{table}


We have batched Tor Browser releases according to what new, unblocked
bridges they contained.
For example, all the releases in the 5.0.5/5.0.6/5.5a5 batch contained
the same new bridge, ndnop3:24215,
while the releases in the 5.5/6.0a1 batch appeared after ndnop3:24215 was blocked
but added six fresh bridges.
There are five of these release batches
(we omitted releases that did not have any new bridges).

Blocking of new bridges was delayed, but abrupt.
When a batch contained more than one bridge,
all were blocked at once (within our probing period of 20 minutes).
Across the five batches, we observed blocking delays of
7, 2, 18, 11, and 36 days after the first public release,
and up to 57 days
after the filing of the first ticket,
when bridges were potentially first discoverable.
This fact suggests, to us, that new default bridges
are loaded into the firewall in groups,
and are not, for example, detected and blocked one at a time.

We found that blocking in China was keyed on both IP address and port,
consistent with an observation of Winter and Lindskog in 2012~\cite{Winter2012a}.
For example, many of the bridges happened to have port 22 (SSH) open,
and it remained accessible even as other ports on the same IP address were blocked.
(See riemann:22 in Figure~\ref{fig:timelines} for an example.)
Per-port blocking is what enabled us to run multiple bridges
on the same IP address.

We never saw a case of a bridge being blocked
before a public release of Tor Browser,
despite their being potentially discoverable
at an earlier stage.
The four bridges that were included only in Orbot,
not in Tor Browser, were never blocked.
From these facts, we infer
that the censors in China probably learn of bridges
not from the bug tracker (which would have revealed Orbot's bridges),
nor from source code inspection (which might have gotten the bridges blocked before release),
but only from public releases.
The 5.5.1/5.5.2/6.0a2 batch is an interesting case because
there was an unusually large gap (about 10 days) between stable and alpha releases:
there were stable releases 18 and 11 days,
and an alpha release 8 days, before
JonbesheSabz:80 and Azadi:443 were blocked.
The bridge ndnop4:27668 did not appear in a release,
but only in BridgeDB, and was not blocked.

\begin{figure}
\includegraphics{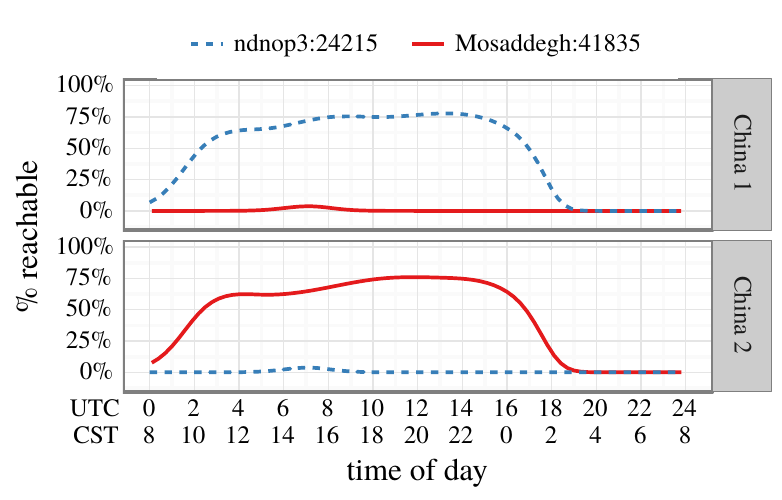}
\caption{
Rates of reachability by time of day
for two bridges from two sites,
between February~1 and March~15, 2016.
There is a diurnal blocking pattern
in both China sites,
though not the same bridges are affected at both sites.
China Standard Time (CST) is UTC+08:00.
}
\label{fig:diurnal}
\end{figure}

There is a conspicuous on--off pattern
in the reachability of certain bridges from China,
for example ndnop3:24215 in China~1
and Mosaddegh:41835 in China~2.
The pattern is roughly periodic with a period of 24~hours.
Figure~\ref{fig:diurnal} averages many 24-hours periods
to show the reachability against time of day of two bridges.
The presence of the diurnal pattern appears to depend on both the bridge and the probing site,
perhaps depending on the network path,
as the same bridges do not show the pattern at both sites.
The pattern can come and go, for example in riemann:443 in China~1.

The China sites also display what are apparently
temporary failures of censorship, stretches of a few hours
during which otherwise blocked bridges were reachable.
Intriguingly, one of these corresponds to a known
failure of the Great Firewall that was documented in the press~\cite{scmp-gfw}.
On March~27, Google services---usually blocked in China---were reachable from
about 15:30 to 17:15 UTC.
This time period is a subset of one in which our bridges
were reachable,
which went from about 10:00 to 18:00 UTC on that day.



\begin{table}
\setlength{\tabcolsep}{3pt}
\begin{tabular}{l l l l l}
\textbf{Batch} & \textbf{Date} & \textbf{Day} & \textbf{Time} \\
\hline
5.0.5/5.0.6/5.5a5 & 22~Dec & Tue & 09:00 {\scriptsize UTC} / 17:00 {\scriptsize CST} \\
5.5/6.0a1         & 29~Jan & Fri & 06:00 {\scriptsize UTC} / 14:00 {\scriptsize CST} \\
5.5.1/5.5.2/6.0a2 & 23~Feb & Tue & 02:40 {\scriptsize UTC} / 10:40 {\scriptsize CST} \\
5.5.4/6.0a4       & 29~Mar & Tue & 06:00 {\scriptsize UTC} / 14:00 {\scriptsize CST} \\
5.5.5/6.0a5/6.0   & 01~Jun & Wed & 02:40 {\scriptsize UTC} / 10:40 {\scriptsize CST} \\
\end{tabular}
\caption{
Day and time of blocking events in China.
Times signify the end of an initial period
of continual or near-continual reachability
of all bridges in a batch of releases.
All bridges in a batch ceased their initial reachability
within 20~minutes after the time shown.
}
\label{tab:blocking}
\end{table}

We wondered whether blocking always occurs on the same
day of the week.
It turns out not to be the case,
as our five blocking events happened on
Tuesday~($\times 3$),
Wednesday~($\times 1$),
and Friday~($\times 1$)---see Table~\ref{tab:blocking}.
However, there may be a pattern in the time of day.
In two cases, the last successful probe happened within the 20~minutes following
02:40~UTC,
and in another two cases,
it was just after 06:00~UTC,
The remaining event happened at 09:00~UTC.
It is hard to make inferences from these limited data,
but they, along with the variable delay in blocking,
suggest a blocking procedure that is part manual and part automatic:
a manual process discovers bridges after an unpredictable delay;
then a periodic, automatic process causes the blocks to take effect.

\section{Further questions}


The data present interesting questions
that call for additional experiments.
Orbot's bridges were blocked only as a side effect
of their being blocked in Tor Browser---bridges
exclusive to Orbot remained reachable.
This suggests a further test
to see whether the censor treats stable and alpha releases differently:
include different sets of bridges in each,
and see whether both or only one of the sets gets blocked.
(This idea has the privacy disadvantage that a network eavesdropper
could infer whether someone is running a stable or an alpha release
by watching the IP addresses they connect to.)

Once in possession of a software release,
how does the censor extract the bridge addresses?
There are several possibilities.
They may have a program automatically parse the file containing bridges;
or a person may have to read the file and enter the bridge addresses manually.
They could simply run the browser in a black-box fashion
and note what addresses it connects to.
In order to distinguish these cases,
one could include commented-out or invalid
bridge addresses in the list.

Table~\ref{tab:blocking} shows that our bridges
got blocked only on weekdays during daylight hours
(China Standard Time).
The five blocking events occurred at three
apparently discrete times of day.
It will be interesting to observe more blocking events
and see whether these patterns continue to hold.

We were surprised not to find any blocking of the default bridges in Iran.
The censorship system in Iran has been documented
to effect blocking through means such as bandwidth throttling~\cite{Anderson2013a,Aryan2013a}
and blocking the IP addresses of the Tor directory authorities~\cite{torbug-12727}.
Throttling would affect obfs4 users but
blocking the directory authorities would not,
as bridges serve Tor directory information in-band.
Iran's censors do not seem to rely on blocking the default bridges,
at least those we tested.



It would be useful to augment our reachability tests with
traceroutes, so we could see where all the routers are between the probe site
and the bridge, and where packets are dropped when bridges are blocked.
Differences in routing might explain why some bridges
were periodically reachable from one China site but not from the other.

The fact that proxy blocking delay is on the order of days
suggests an obvious circumvention strategy:
if the censor blocks new bridges after $n$ days,
introduce new ones every $n-1$ days.
This could be accomplished through more frequent releases,
or obfs4 could be modified to try different destination ports
according to some schedule that depends on the time since release.
The interesting question then becomes the meta-delay:
how long does the censor take to catch on to this new pattern?

\section{Acknowledgments}

We express our thanks to those who helped us by running bridges
or providing us with probe sites, including
Percy Alpha,
Nima Fatemi,
Linus Nordberg,
Henry de~Valence,
and others who remain anonymous.
We thank the developers of Tor Browser and Orbot
for their timeliness in adding new bridges.
We appreciate our discussions with censorship researchers at
the University of California, Berkeley and
the International Computer Science Institute.

\section{Availability}

Our code and data are available from the supporting web page:
\url{https://www.bamsoftware.com/proxy-probe/}.

{\footnotesize \bibliographystyle{acm}
\bibliography{proxy-probe}}

\end{document}